\documentclass[aps,showkeys,twocolumn,preprintnumbers,amsmath,amssymb,superscriptaddress,floatfix,nofootinbib]{revtex4}
\usepackage{graphicx,color,dcolumn,booktabs,bm}
\usepackage{amsmath}
\usepackage{graphicx}
\usepackage{longtable,lscape}
\usepackage{txfonts}
\usepackage{mathrsfs}
\usepackage{IEEEtrantools}
\usepackage{overpic}
\usepackage{epsfig}
\usepackage{amssymb}
\usepackage{bbding}
\usepackage{rotating}
\usepackage{epstopdf}
\usepackage{appendix}
\usepackage{indentfirst}
\usepackage{feynmf}   
\usepackage{slashed}  
\usepackage{cases}
\usepackage{color}
\usepackage{ulem}
\usepackage{multirow}
\usepackage{graphicx,color,dcolumn,booktabs,bm}
\usepackage{cases}
\usepackage{array}
\usepackage{enumerate}

\graphicspath{{Figures/}} %

\usepackage[colorlinks, citecolor=blue,anchorcolor=red,menucolor=red, linkcolor=red,filecolor=red,runcolor=red,urlcolor=blue,frenchlinks=red]{hyperref}

\begin{document}
\title{A coupled-channel perspective analysis on bottom-strange molecular pentaquarks}
\author{Qing-Fu Song}\email{242201003@csu.edu.cn}

\affiliation{School of Physics, Central South University, Changsha 410083, China}

\author{Qi-Fang L\"{u}}\email{lvqifang@hunnu.edu.cn}

\affiliation{Department of Physics, Hunan Normal University, Changsha 410081, China}
\affiliation{Key Laboratory of Low-Dimensional Quantum Structures and Quantum Control of Ministry of Education, Changsha 410081, China}
\affiliation{Key Laboratory for Matter Microstructure and Function of Hunan Province, Hunan Normal University, Changsha 410081, China}

\author{Xiaonu Xiong}\email{xnxiong@csu.edu.cn}

\affiliation{School of Physics, Central South University, Changsha 410083, China}

\begin{abstract}

In the present work, we systematically study various bottom-strange molecular pentaquarks to search for possible bound states and resonances by adopting a one boson exchange model within the complex scaling method. According to our calculations, we predict several bound and resonant states for bottom baryon $Y_{b}(\Lambda_b,\Sigma_b) \bar K^{(*)}$ and $Y_{b} K^{(*)}$ systems. In particular, a bound state in the $I(J^P)=1/2(1/2^-)\Sigma_b \bar{K}/\Lambda_b \bar{K}^*/\Sigma_b \bar{K}^*$ system may correspond to the particle $\Xi_b(6227)$. Meanwhile, the predicted bound state with mass in the range of $6303\sim6269~\rm{MeV}$ in the  $I(J^P)=1/2(1/2^-)\Sigma_bK/\Lambda_bK^*/\Sigma_bK^*$ system is flavor exotic and does not appear in the spectroscopy of conventional baryons, which may help clarify the nature of $\Xi_b(6227)$. We hope that our study can provide useful guidance for future experimental searches.

\end{abstract}


\maketitle
\section{introduction}
After the discovery of the $X(3872)$ {by} the Belle experiment~\cite{Belle:2003nnu} in 2003, a large amount of data has been accumulated {over} the past two decades in high energy collision experiments. {Meanwhile}, a series of new {phenomenological} studies related to the $XYZ$ and $P_c/T_{cc}$ states have been reported~\cite{Liu:2019zoy,Liu:2015fea}.  
A {detailed} investigation of those exotic hadron states provides new insights {into} decoding their internal structures, which may deepen our understanding of the nonperturbative properties of quantum chromodynamics (QCD). Since many new particles {are located} near the hadron-hadron thresholds, these states could be naturally interpreted as candidates {for} molecular states~\cite{Chen:2022asf,Dong:2017gaw,Guo:2017jvc,Karliner:2017qhf,Zou:2021sha,Mai:2022eur,Zou:2013af}. It is highly desirable to identify those {molecular} states out of {the numerous} candidates and predict {more of them} for experimental searches, which will motivate the experimental search {for} such molecular states.

In 2021, the LHCb Collaboration reported two resonances, namely $X_0(2900)$ and $X_1(2900)$, in the $D^-K^{+}$ invariant mass spectrum by analyzing the decay amplitude of the $B^+ \to D^+ D^- K^+$ decay channel~\cite{LHCb:2020pxc,LHCb:2020bls}. Since these two states are located near the $\bar{D}^{*}K^{*}$ and $\bar{D}_{1}^{*}K^{*}$ {thresholds}, they are regarded as hadronic {molecule} candidates~\cite{Chen:2020aos,He:2020btl,Burns:2020epm,Agaev:2020nrc,Xiao:2020ltm,Kong:2021ohg,Ke:2022ocs}. Recently, in the analysis of the $D_{s}^{+}\pi^{+}$ and $D_{s}^{+}\pi^{-}$ invariant mass {spectra}, the LHCb Collaboration has observed two new peaks, $T_{c\bar{s}0}^{0}(2900)$ and $T_{c\bar{s}0}^{++}(2900)$, whose masses and widths are $M(T_{c\bar{s}0}^{0})=2892\pm14\pm15$ $\rm MeV$, $\Gamma=119\pm26\pm12$ $\rm MeV$ and $M(T_{c\bar{s}0}^{++})=2921\pm17\pm19$ $\rm MeV$, $\Gamma=137\pm32\pm14$ $\rm MeV$~\cite{LHCb:2022sfr,LHCb:2022lzp}, respectively. Given their near-threshold behaviors and quantum numbers, these two $T_{c\bar{s}0}^{0(++)}$ states are proposed as isovector $D^*K^*$ molecules with $J^{P}=0^+$~\cite{Agaev:2022eyk,Wang:2023hpp,An:2022vtg}.

Until now, most molecular candidates were observed in the charm sector, while the experimental observations in the bottom sector are still scarce. In 2006, the DØ Collaboration announced a narrow structure, referred to as the $X(5568)$, in the $B_{s}^{0}\pi^{\pm}$ channel~\cite{D0:2016mwd}. Then, the LHCb Collaboration investigated the $B_{s}^{0}\pi^{\pm}$ invariant mass spectrum, but no significant signal {was} found~\cite{LHCb:2016dxl}. Later, the ATLAS, CDF, and CMS Collaborations~\cite{ATLAS:2018udc,CDF:2017dwr,CMS:2017hfy} released similar results. Meanwhile, the $X(5568)$ has been theoretically discussed in previous works~\cite{Chen:2016ypj,Xiao:2016mho,Agaev:2016urs,Guo:2016nhb,Albaladejo:2016eps,Albaladejo:2016lbb}, and {cannot} be assigned as an isovector $BK$ molecular state. In 2021, the LHCb Collaboration reported two states in the $B^{\pm}K^{\pm}$ mass spectrum, which are {referred to} as $B_{sJ}(6063)$ and $B_{sJ}(6114)$. If the missing {photon} from the $B^{*\pm} \to B^{\pm} \gamma$ was taken into consideration, the masses and widths were measured to be $B_{sJ}(6109): M = 6108.8 \pm 1.1 \pm 0.7~\rm MeV$ \text{ and } $\Gamma = 22 \pm 5 \pm 4~\rm MeV$, and $B_{sJ}(6158): M = 6158 \pm 4 \pm 5~\rm MeV$ \text{ and } $\Gamma = 72 \pm 18 \pm 25~\rm MeV$~\cite{LHCb:2020pet}, respectively. In theory, the $B_{sJ}(6158)$ {has been} widely investigated in the literature~\cite{Kong:2021ohg}. Some of the existing works suggested that $B_{sJ}(6158)$ can be interpreted as a $\bar{B}K^*$ molecular state with $I(J^{PC}) = 0(1^+)$. Also, several works showed that the existence of $\bar{B}^{(*)}K^{(*)}(B^{(*)}\bar{K}^{(*)})$ molecular states {is} allowed~\cite{Kolomeitsev:2003ac,Guo:2006fu,Guo:2006rp,Sun:2018zqs}.

\begin{figure}
    \centering
    \includegraphics[width=1\linewidth]{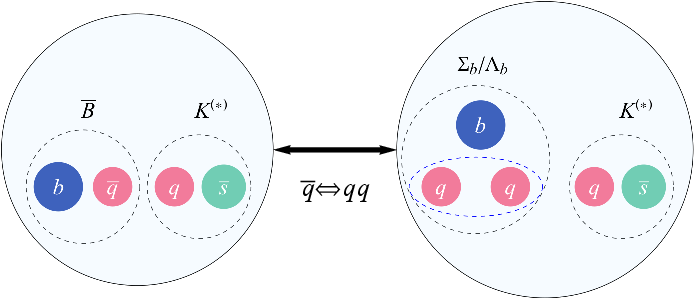}
    \caption{ A sketch of heavy superflavor symmetry between $\Sigma_b(\Lambda_b)K^{(*)}$  pentaquarks and $\bar{B}K^{(*)}$ tetraquarks. $q$ stands for the light quarks ($u$ or $d$).}
    \label{heavy}
\end{figure}

It can be seen that numerous exotic hadronic molecular states containing heavy quarks have been observed experimentally. The heavy quark symmetry is supposed to have been proven to play a significant role in predicting undiscovered states and understanding their production mechanisms, {which has stimulated} several theoretical studies~\cite{Asanuma:2023atv,Tanaka:2024siw,Wang:2023eng,Sakai:2023syt}. In a previous work, {the authors investigated the open-charm molecular counterparts of the newly observed} $T_{c\bar{s}}^{a0(++)}$ {composed of $Y_{c}(\Lambda_c, \Sigma_c)$ baryons and strange mesons $K^{(*)}$, by adopting the one-boson exchange model}. From their estimations, {some bound states may exist that correspond to the newly observed} $T_{c\bar{s}}^{a0(++)}$~\cite{Chen:2022svh}. According to heavy quark symmetry, {in the bottom sector}, the light diquark in heavy baryons $\Sigma_b/\Lambda_b$ has the same color structure as a light antiquark $\bar{q}$, as shown in Figure~\ref{heavy}. If the $B_{sJ}(6158)$ can be explained as a $\bar{B}K^*$ molecular state with $I(J^{PC})=0(1^+)$, {one should also expect the existence of isoscalar} $\bar{B}^{(*)}K^{(*)}$ molecular state. {Under these circumstances}, it is natural to conjecture whether {open-bottom molecular pentaquarks could also exist}. Moreover, it is worth mentioning that, in 2018, the LHCb {Collaboration reported a peak in both the} $\Lambda_{b}^{0}K^{-}$ and $\Xi_{b}^{0}\pi^{-}$ invariant mass spectra, named $\Xi_{b}(6227)$~\cite{LHCb:2018vuc}. However, until now, whether the $\Xi_{b}(6227)$ should be accommodated into traditional baryon $\lambda$-mode $P$-wave $\Xi_{b}^{\prime}$ baryon with $J^{P}=3/2^-$ or $5/2^-$~\cite{Chen:2018orb,He:2021xrh, Wang:2018fjm,Cui:2019dzj}, or {instead as a pure molecular state $\Sigma_{b}\bar{K}$ with $J^P=1/2^-$}~\cite{Huang:2018bed,Zhu:2020lza,Mutuk:2024elj,Ozdem:2021vry}is still on discussion. Thus, it is {both urgent and necessary to explore the possibility of} $\Xi_b(6227)$ being a molecular state and to {predict more bottom-strange molecular pentaquark candidates for future experimental searches}.

Recently, we systematically {studied} the hidden bottom molecular tetraquark with complex scaling method by adopting one-boson-exchange (OBE) model~\cite{Song:2024ngu}. {In the present work, utilizing the same {formalism}, we systematically study various bottom-strange molecular pentaquarks to search for possible bound states and resonances by adopting {the} complex scaling method~\cite{Aguilar:1971ve,Balslev:1971vb,Moiseyev:1998gjp,Ho:1983lwa} and Gaussian expansion method~\cite{Hiyama:2003cu,Hiyama:2018ivm}.}  For bottom baryon $Y_{b}(\Lambda_b,\Sigma_b)$ and anti-strange meson $\bar{K}^{(*)}$ interactions, our calculations demonstrate that some bound and resonant states are {revealed}. For instance, in the $I(J^P)=1/2(1/2^-)$  $\Sigma_{b}\bar{K}/\Lambda_{b}\bar{K}^*/\Sigma_{b}\bar{K}^*$ system, we obtain a bound state below $\Sigma_{b}\bar{K}$ threshold that can be regarded as the particle $\Xi_{b}(6227)$. Meanwhile, we extend our study to $Y_{b}K^{(*)}$ systems, and find two flavor exotic bound states, which can be searched {for} in future experiments.

The rest of this paper is organized as follows. We briefly introduce the formalism of effective interactions and complex scaling method in Sec.~\ref{model}. In Sec.~\ref{results}, we present the numerical results and discussions for the $Y_{b}$$K^{(*)}$ and $Y_{b}$$\bar{K}^{(*)}$  systems. Finally, we summarize in Sec.~\ref{sum}.

\section{Formalism of effective interaction and complex scaling method}\label{model}
\subsection{The effective interactions}
In this work, we adopt the one-boson-exchange model to describe the {interactions} between the hadrons and analyze the formation mechanisms of molecular states. The {chiral} symmetric interacting Lagrangian, which corresponds to the coupling between a bottom baryon and {a light meson}, can be constructed as~\cite{Liu:2011xc}

\begin{eqnarray}
\mathcal{L}_{\mathcal{B}_{\bar{3}}} &=& l_B\langle\bar{\mathcal{B}}_{\bar{3}}\sigma\mathcal{B}_{\bar{3}}\rangle
          +i\beta_B\langle\bar{\mathcal{B}}_{\bar{3}}v^{\mu}(\mathcal{V}_{\mu}-\rho_{\mu})\mathcal{B}_{\bar{3}}\rangle,\label{lag1}\\
\mathcal{L}_{\mathcal{B}_{6}} &=&  l_S\langle\bar{\mathcal{S}}_{\mu}\sigma\mathcal{S}^{\mu}\rangle
         -\frac{3}{2}g_1\varepsilon^{\mu\nu\lambda\kappa}v_{\kappa}
         \langle\bar{\mathcal{S}}_{\mu}A_{\nu}\mathcal{S}_{\lambda}\rangle\nonumber\\
    &+&i\beta_{S}\langle\bar{\mathcal{S}}_{\mu}v_{\alpha}
    \left(\mathcal{V}_{ab}^{\alpha}-\rho_{ab}^{\alpha}\right) \mathcal{S}^{\mu}\rangle
    +\lambda_S\langle\bar{\mathcal{S}}_{\mu}F^{\mu\nu}(\rho)\mathcal{S}_{\nu}\rangle,~\\
\mathcal{L}_{\mathcal{B}_{\bar{3}}\mathcal{B}_6} &=& ig_4\langle\bar{\mathcal{S}^{\mu}}A_{\mu}\mathcal{B}_{\bar{3}}\rangle
         +i\lambda_I\varepsilon^{\mu\nu\lambda\kappa}v_{\mu}\langle \bar{\mathcal{S}}_{\nu}F_{\lambda\kappa}\mathcal{B}_{\bar{3}}\rangle+h.c..\label{lag2}
\end{eqnarray}
The axial current $A_{\mu}$, {the} vector current $\mathcal{V}_{\mu}$, and the vector meson field strength tensor $F^{\mu\nu}(\rho)$ are defined {as follows}

\begin{alignat}{2}
    \mathcal{V}_{\mu} &= &&\frac{1}{2}(\xi^{\dag}\partial_{\mu}\xi+\xi\partial_{\mu}\xi^{\dag}),\\
A_{\mu} &= &&\frac{1}{2}(\xi^{\dag}\partial_{\mu}\xi-\xi\partial_{\mu}\xi^{\dag}),\\
F_{\mu\nu}(\rho)&=&&\partial_{\mu}\rho_{\nu}-\partial_{\nu}\rho_{\mu}+[\rho_{\mu},\rho_{\nu}],
\end{alignat}
respectively.  
Here, $\xi=\text{exp}({P}/f_{\pi})$ and $\rho_{ba}^{\mu}=ig_V{V}_{ba}^{\mu}/\sqrt{2}$. The $\mathcal{B}_{\bar{3}}$, {$\mathcal{B}_{\mu} = -\sqrt{\frac{1}{3}}(\gamma_\mu + \nu_\mu)\gamma^5 \mathcal{B}_6 + \mathcal{B}_{6\mu}^*$}, ${P}$, and ${V}$ denote the matrices of the ground {states} of singly heavy baryon multiplets in $\bar{3}_{F}$, $6_{F}$, light pseudoscalar, and vector mesons, respectively, whose explicit {forms} read

{
\begin{eqnarray}
\mathcal{B}_{6} &=& 
\begin{pmatrix}
\Sigma_b^{+} & \frac{\Sigma_b^0}{\sqrt{2}}&\frac{\Xi_b^{(',*)0}}{\sqrt{2}} \\
\frac{\Sigma_b^0}{\sqrt{2}} & \Sigma_b^{-}&\frac{\Xi_b^{(',*)-}}{\sqrt{2}}\\
\frac{\Xi_b^{(',*)0}}{\sqrt{2}}&\frac{\Xi_b^{(',*)-}}{\sqrt{2}}&\Omega_b^{(*)-}
\end{pmatrix}, \quad
\mathcal{B}_{\bar{3}} = 
\begin{pmatrix}
0 & \Lambda_b^0 &\Xi_b^0\\
-\Lambda_b^0 & 0&\Xi_b^-\\
-\Xi_b^0&-\Xi_b^-&0\\
\end{pmatrix},
\end{eqnarray}
\begin{eqnarray}
P &=& 
\begin{pmatrix}
\frac{\pi^0}{\sqrt{2}}+\frac{\eta}{\sqrt{6}} & \pi^+&K^+ \\
\pi^- & -\frac{\pi^0}{\sqrt{2}}+\frac{\eta}{\sqrt{6}}&K^0\\
K^-&\bar{K}^0&\frac{2}{\sqrt{6}}\eta\\
\end{pmatrix}, \quad
\end{eqnarray}
\begin{eqnarray}
V = 
\begin{pmatrix}
\frac{\rho^0}{\sqrt{2}}+\frac{\omega}{\sqrt{2}} & \rho^+ &K^{*+}\\
\rho^- & -\frac{\rho^0}{\sqrt{2}}+\frac{\omega}{\sqrt{2}}&K^{*0}\\
K^{*-}&\bar{K}^{*0}&\phi\\
\end{pmatrix}.
\end{eqnarray}}
Under the SU(3) symmetry, the effective Lagrangians describing the interactions between {the strange mesons} and {light mesons} can be expressed as~\cite{Lin:1999ad}
\begin{eqnarray}
\mathcal{L}_{PPV} &=& \frac{ig}{2\sqrt{2}}\langle\partial^{\mu}P\left(PV_{\mu}-V_{\mu}P\right\rangle, \label{lag4}\\
\mathcal{L}_{VVP} &=& \frac{g_{VVP}}{\sqrt{2}}\epsilon^{\mu\nu\alpha\beta}
       \left\langle\partial_{\mu}V_{\nu}\partial_{\alpha}V_{\beta}P\right\rangle,\label{lag5}\\
\mathcal{L}_{VVV} &=& \frac{ig}{2\sqrt{2}}\langle\partial^{\mu}V^{\nu}\left(V_{\mu}V_{\nu}-V_{\nu}V_{\mu}\right)\rangle. \label{lag6}
\end{eqnarray}
More specifically, one can further write the effective Lagrangian depicting the couplings as
\begin{eqnarray}
\mathcal{L}_{\sigma} &=& l_B\langle \bar{\mathcal{B}}_{\bar{3}}\sigma\mathcal{B}_{\bar{3}}\rangle
      -l_S\langle\bar{\mathcal{B}}_6\sigma\mathcal{B}_6\rangle,\\
\mathcal{L}_{{P}} &=&
        i\frac{g_1}{2f_{\pi}}\varepsilon^{\mu\nu\lambda\kappa}v_{\kappa}\langle\bar{\mathcal{B}}_6
        \gamma_{\mu}\gamma_{\lambda}\partial_{\nu}{P}\mathcal{B}_6\rangle\nonumber\\
      &-&\sqrt{\frac{1}{3}}\frac{g_4}{f_{\pi}}\langle\bar{\mathcal{B}}_6\gamma^5
      \left(\gamma^{\mu}+v^{\mu}\right)\partial_{\mu}{P}\mathcal{B}_{\bar{3}}\rangle+h.c.,\\
\mathcal{L}_{{V}} &=& \frac{1}{\sqrt{2}}\beta_Bg_V\langle\bar{\mathcal{B}}_{\bar{3}}v\cdot{V}\mathcal{B}_{\bar{3}}\rangle
   -\frac{\beta_Sg_V}{\sqrt{2}}\langle\bar{\mathcal{B}}_6v\cdot{V}\mathcal{B}_6\rangle\nonumber\\
    &-&\frac{\lambda_Ig_V}{\sqrt{6}}\varepsilon^{\mu\nu\lambda\kappa}v_{\mu}\langle \bar{\mathcal{B}}_6\gamma^5\gamma_{\nu}
        \left(\partial_{\lambda} {V}_{\kappa}-\partial_{\kappa} {V}_{\lambda}\right)\mathcal{B}_{\bar{3}}\rangle+h.c.\nonumber\\
        &-&i\frac{\lambda g_V}{3\sqrt{2}}\langle\bar{\mathcal{B}}_6\gamma_{\mu}\gamma_{\nu}
    \left(\partial^{\mu} {V}^{\nu}-\partial^{\nu} {V}^{\mu}\right)
    \mathcal{B}_6\rangle,\\
\mathcal{L}_{K^{(*)}K^{(*)}\sigma} &=& g_{\sigma }m_K\bar{K} K\sigma-g_{\sigma }m_{K^*}\bar{K}^{*}\cdot K^{*}\sigma,\\
\mathcal{L}_{P KK^*} &=& \frac{ig}{4}\left[\left(\bar{K}^{*\mu} K-\bar{K} K^{*\mu}\right)\left(\bm{\tau}\cdot\partial_{\mu}\bm{\pi}+\frac{\partial_{\mu}{\eta}}{\sqrt{3}}\right)\right.\nonumber\\
   &+&\left.\left(\partial_{\mu}\bar{K} K^{*\mu}-\bar{K}^{*\mu}\partial_{\mu}K\right)\left(\bm{\tau}\cdot\bm{\pi}+\frac{\eta}{\sqrt{3}}\right)\right],\\
\mathcal{L}_{{V} KK} &=& \frac{ig}{4}\left[\bar{K}\partial_{\mu}K
       -\partial_{\mu}\bar{K}K\right]\left(\bm{\tau}\cdot\bm{\rho}^{\mu}+{\omega}^{\mu}\right),\\
\mathcal{L}_{{V} K^*K^*} &=& \frac{ig}{4}
       \left[\left(\bar{K}_{\mu}^*\partial^{\mu}K^{*\nu}-\partial^{\mu}\bar{K}^{*\nu} K_{\mu}^*\right)\left(\bm{\tau}\cdot\bm{\rho}_{\nu}+\omega_{\nu}\right)\right.\nonumber\\
   &&\left.+\left(\partial^{\mu}\bar{K}^{*\nu}K_{\nu}^*-\bar{K}_{\nu}^*\partial^{\mu}K^{*\nu}\right)
       \left(\bm{\tau}\cdot\bm{\rho}_{\mu}+\omega_{\mu}\right)\right.\nonumber\\
       &&\left.+\left(\bar{K}_{\nu}^* K^*_{\mu}-\bar{K}_{\mu}^*K^*_{\nu}\right)
       \left(\bm{\tau}\cdot\partial^{\mu}\bm{\rho}^{\nu}+\partial^{\mu}\omega^{\nu}\right)\right],\\
\mathcal{L}_{P K^*K^*} &=& g_{VVP}\varepsilon_{\mu\nu\alpha\beta}
     \partial^{\mu}\bar{K}^{*\nu}\partial^{\alpha}K^{*\beta}\left(\bm{\tau}\cdot\bm{\pi}+\frac{\eta}{\sqrt{3}}\right),\\
\mathcal{L}_{V KK^*} &=& g_{VVP}\varepsilon_{\mu\nu\alpha\beta}
     \left(\partial^{\mu}\bar{K}^{*\nu}K+\bar{K}\partial^{\mu}{K}^{*\nu}\right)\nonumber\\
   &&\left(\bm{\tau}\cdot\partial^{\alpha}\bm{\rho}^{\beta}+\partial^{\alpha}{\omega}^{\beta}\right).
\end{eqnarray}

\begin{table*}[!t]
	\renewcommand\arraystretch{1.6}
	\caption{\label{potential}  The effective  potentials for $Y_b(\Lambda_b,\Sigma_b)\bar{K}^{(*)}$ systems.}
	\begin{ruledtabular}
		\begin{tabular}{ccccccccc}
			&Processes
			&Effective potentials\\\hline
      &\multirow{1}{*}{$V_{\Lambda_b\bar{K}^*\to\Lambda_b\bar{K}^*} $}&$l_Bg_{\sigma}(\bm{\epsilon}_2\cdot\bm{\epsilon}_4^{\dag})\chi_3^{\dag}\chi_1Y(\Lambda,m_{\sigma},r)-\frac{\beta_Bg_Vg}{4}(\bm{\epsilon}_2\cdot\bm{\epsilon}_4^{\dag})\chi_3^{\dag}\chi_1Y(\Lambda,m_{\omega},r)$\\\hline
    &\multirow{1}{*}{$V_{\Lambda_b\bar{K}^*\to\Sigma_b\bar{K}^*}$}&$-\frac{1}{6}\frac{g_4g_{VVP}}{f_{\pi}}
\mathcal{F}_1(r,\bm{\sigma},i\bm{\epsilon}_2\times\bm{\epsilon}_4^{\dag})
     Y(\Lambda_0,m_{\pi0},r)-\frac{1}{6\sqrt{2}}\frac{\lambda_Ig_Vg}{m_{K^*}}
     \mathcal{F}_2(r,\bm{\sigma},i\bm{\epsilon}_2\times\bm{\epsilon}_4^{\dag})
     Y(\Lambda_0,m_{\rho0},r)$\\\hline
			&\multirow{3}{*}{$V_{\Sigma_b\bar{K}^*\to\Sigma_b\bar{K}^*}$} &$ +\frac{1}{2}l_Sg_{\sigma}\chi_3^{\dag}\chi_1
\bm{\epsilon}_2\cdot\bm{\epsilon}_3^{\dag}Y(\Lambda,m_{\sigma},r)+\frac{g_1g_{VVP}}{6\sqrt{2}f_{\pi}}
\mathcal{F}_1(r,\bm{\sigma},i\bm{\epsilon}_2\times\bm{\epsilon}_4^{\dag})
     \mathcal{G}(I)Y(\Lambda,m_{\pi},r)-\frac{g_1g_{VVP}}{18\sqrt{2}f_{\pi}}
\mathcal{F}_1(r,\bm{\sigma},i\bm{\epsilon}_2\times\bm{\epsilon}_4^{\dag})
     Y(\Lambda,m_{\eta},r)$&\\
		&&$+\frac{1}{8}\beta_Sg_Vg\chi_3^{\dag}\chi_1
     \bm{\epsilon}_2\cdot\bm{\epsilon}_3^{\dag}\mathcal{G}(I)Y(\Lambda,m_{\rho},r)+\frac{\lambda_Sg_Vg}{8\sqrt{3}m_{\Sigma_b}}\chi_3^{\dag}\chi_1
     \bm{\epsilon}_2\cdot\bm{\epsilon}_3^{\dag}\mathcal{G}(I)\nabla^2Y(\Lambda,m_{\rho},r)-\frac{\lambda_Sg_Vg}{24\sqrt{3}m_{K^*}}
\mathcal{F}_2(r,\bm{\sigma},i\bm{\epsilon}_2\times\bm{\epsilon}_4^{\dag})
     \mathcal{G}(I)Y(\Lambda,m_{\rho},r)$&\\
     &&$-\frac{1}{8}\beta_Sg_Vg\chi_3^{\dag}\chi_1
     \bm{\epsilon}_2\cdot\bm{\epsilon}_3^{\dag}Y(\Lambda,m_{\omega},r)-\frac{\lambda_Sg_Vg}{8\sqrt{3}m_{\Sigma_b}}\chi_3^{\dag}\chi_1
     \bm{\epsilon}_2\cdot\bm{\epsilon}_3^{\dag}\nabla^2Y(\Lambda,m_{\omega},r)+\frac{\lambda_Sg_Vg}{24\sqrt{3}m_{K^*}}
\mathcal{F}_2(r,\bm{\sigma},i\bm{\epsilon}_2\times\bm{\epsilon}_4^{\dag})
     Y(\Lambda,m_{\omega},r)$&\\\hline
      &\multirow{2}{*}{$V_{\Sigma_b\bar{K}\to\Sigma_b\bar{K}}$}&$\frac{1}{2}l_Sg_{\sigma}\chi_3^{\dag}\chi_1Y(\Lambda,m_{\sigma},r)+\frac{\mathcal{G}(I)}{8}\beta_Sg_Vg\chi_3^{\dag}\chi_1Y(\Lambda,m_{\rho},r)-\frac{\mathcal{G}(I)}{24m_{\Sigma_b}}\lambda_Sg_Vg\chi_3^{\dag}\chi_1
    \nabla^2Y(\Lambda,m_{\rho},r)$\\
    &&$-\frac{1}{8}\beta_Sg_Vg\chi_3^{\dag}\chi_1Y(\Lambda,m_{\omega},r)+\frac{1}{24m_{\Sigma_b}}\lambda_Sg_Vg\chi_3^{\dag}\chi_1\nabla^2Y(\Lambda,m_{\omega},r)$&\\\hline
			&\multirow{1}{*}{$V_{\Lambda_b\bar{K}^*\to\Sigma_b\bar{K}} $}&$ \frac{1}{6}\frac{g_4g}{f_{\pi}\sqrt{m_Km_{K^*}}}
    \mathcal{F}_1(r,\bm{\sigma},\bm{\epsilon}_2)
     U(\Lambda_1,m_{\pi1},r) -\frac{\lambda_Ig_Vg_{VVP}}{3\sqrt{2}}\sqrt{\frac{m_{K^*}}{m_K}}
    \mathcal{F}_2(r,\bm{\sigma},\bm{\epsilon}_2)
     Y(\Lambda_1,m_{\rho1},r)$\\\hline
     &\multirow{2}{*}{$V_{\Sigma_b\bar{K}^*\to\Sigma_b\bar{K}}$}&$-\frac{g_1g\mathcal{F}_1(r,\bm{\sigma},\bm{\epsilon}_2)}
  {24\sqrt{2}f_{\pi}\sqrt{m_Km_{K^*}}}
     \mathcal{G}(I)Y(\Lambda_2,m_{\pi2},r)+\frac{g_1g}{72\sqrt{2}f_{\pi}\sqrt{m_Km_{K^*}}}
    \mathcal{F}_1(r,\bm{\sigma},\bm{\epsilon}_2)
     Y(\Lambda_2,m_{\eta2},r)$\\
     &&$+\frac{\lambda_Sg_Vg_{VVP}}{6\sqrt{3}}\sqrt{\frac{m_{K^*}}{m_K}}
    \mathcal{F}_2(r,\bm{\sigma},\bm{\epsilon}_2)
     \mathcal{G}(I)Y(\Lambda_2,m_{\rho2},r)-\frac{\lambda_Sg_Vg_{VVP}}{6\sqrt{3}}\sqrt{\frac{m_{K^*}}{m_K}}
    \mathcal{F}_2(r,\bm{\sigma},\bm{\epsilon}_2)
     Y(\Lambda_2,m_{\omega2},r)$&\\
		\end{tabular}
	\end{ruledtabular}
\end{table*}

{
Before applying the potentials in coordinate space, we adopt the Breit approximation, a standard approach within the one-boson-exchange (OBE) framework. In this approximation, the relativistic scattering amplitude is reduced to an instantaneous effective potential in momentum space, which is valid in the low-energy regime with small momentum transfer. This treatment enables us to extract spin-dependent interaction terms, such as central, spin-spin, tensor, and spin-orbit components. For a more detailed discussion, we refer the reader to Refs.~\cite{Breit:1929zz,Breit:1930zza}.
Thus, the effective potential in momentum space reads}

\begin{eqnarray}\label{breit}
\mathcal{V}^{h_{1}h_{2}\to h_{3}h_{4}}(\bm{q}) &=&
          -\frac{\mathcal{M}(h_{1}h_{2}\to h_{3}h_{4})}
          {4\sqrt{m_{1}m_{2}m_{3}m_{4}}},
\end{eqnarray}
in which $\mathcal{M}(h_{1}h_{2}\to h_{3}h_{4})$ denotes the scattering amplitude for the $h_{1}h_{2}\to h_{3}h_{4}$ process{,} and $m_{i}$ is the mass of the particle $h_i$. The Fourier {transform} with respect to $\bm{q}$ leads to the effective potential in position space,

\begin{eqnarray}
\mathcal{V}(\bm{r}) =
          \int\frac{d^3\bm{q}}{(2\pi)^3}e^{i\bm{q}\cdot\bm{r}}
          \mathcal{V}(\bm{q})\mathcal{F}^2(q^2,m_i^2),
\end{eqnarray}
where $\mathcal{F}$ denotes the form factor with explicit form
\begin{eqnarray}\label{FF}
          \mathcal{F}(q^2,m_i^2)= \frac{\Lambda^2-m_i^2}{\Lambda^2-q^2}.
\end{eqnarray}
Here, the parameter $\Lambda$ is introduced as an UV cut-off originates from the fact that the hadrons have a non-zero size to account for the internal structure of the interacting hadrons.

The corresponding one-boson-exchange effective {potentials are} taken from Ref.~\cite{Chen:2022svh} and listed in Table~\ref{potential}, where $\mathcal{G}=-2$ for {the} $I=1/2$ system and $\mathcal{G}=1$ for {the} $I=3/2$ system. The explicit expressions for factors $\mathcal{F}_{1,2}$, $U$ and $Y$ in the
effective potentials listed in Table~\ref{potential} are given

\begin{eqnarray}\nonumber
\mathcal{F}_1(r,\mathbf{a},\mathbf{b})&=&\chi_3^\dagger\bigg(\mathbf{a}\cdot\mathbf{b}\nabla^2+S(\hat{r},\mathbf{a},\mathbf{b})r\frac{\partial}{\partial r}\frac{1}{r}\frac{\partial}{\partial r}\bigg)\chi_1\\\nonumber
\mathcal{F}_2(r,\mathbf{a},\mathbf{b})&=&\chi_3^\dagger\bigg(2\mathbf{a}\cdot\mathbf{b}\nabla^2-S(\hat{r},\mathbf{a},\mathbf{b})r\frac{\partial}{\partial r}\frac{1}{r}\frac{\partial}{\partial r}\bigg)\chi_1\\\nonumber
    U(\Lambda,m,r)&=&\frac{1}{4\pi r}\bigg(\text{cos}(mr)-e^{-\Lambda r}\bigg)-\frac{\Lambda^2+m^2}{8\pi\Lambda}e^{-\Lambda r}\\
    Y(\Lambda,m,r)&=&\frac{1}{4\pi r}(e^{-mr}-e^{-\Lambda r})-\frac{\Lambda^2-m^2}{8\pi\Lambda}e^{-\Lambda r},
\end{eqnarray}
where $S(\hat{r},\mathbf{a},\mathbf{b})\equiv 3(\hat{r}\cdot\mathbf{a})(\hat{r}\cdot\mathbf{b})-\mathbf{a}\cdot\mathbf{b}$. The values of relevant parameters are listed in Table~\ref{parameters}~\cite{Liu:2011xc,Chen:2017xat,Kaymakcalan:1983qq}.
{The subscript labels $m_{\pi 1}$, $m_{\pi 2}$, etc., represent the effective masses of the exchanged mesons, which arise from the energy transfer $q^0$ during the Fourier transformation. Specifically, for the $Y(\Lambda_2, m_{\pi 2}, r)$-type potential, we use
\begin{eqnarray}
m_{\pi2} &=& \sqrt{m_\pi^2 - (q^0)^2}, \quad \Lambda_2 = \sqrt{\Lambda^2 - (q^0)^2}, \\
q^0 &=& \frac{m_1^2 + m_4^2 - m_2^2 - m_3^2}{2(m_3 + m_4)} = \frac{m_{\bar{K}}^2 - m_{\bar{K}^*}^2}{2(m_{\Sigma_b} + m_{\bar{K}})}.
\end{eqnarray}
For the $U(\Lambda_1, m_{\pi 1}, r)$-type potential, the definitions are
\begin{eqnarray}
m_{\pi1} &=& \sqrt{(q^0)^2 - m_\pi^2}, \quad \Lambda_1 = \sqrt{\Lambda^2 - (q^0)^2}, \\
q^0 &=& \frac{m_1^2 + m_4^2 - m_2^2 - m_3^2}{2(m_3 + m_4)} = \frac{m_{\Lambda_b}^2 + m_{\bar{K}}^2 - m_{\bar{K}^*}^2 - m_{\Sigma_b}^2}{2(m_{\Sigma_b} + m_{\bar{K}})}.
\end{eqnarray}
The same conventions apply to other exchanged mesons such as $\rho_{1,2}$, $\eta_2$, and $\omega_{1,2}$.}

\begin{table}[htbp]
\caption{\label{parameters} The relevant parameters adopted in this work.}
\begin{ruledtabular}
\begin{tabular}{ccccccccc}
&Parameters
&Values
\\\hline
&$l_s=2l_B$&7.3\\
&$g_1=(\sqrt{8}/3)g_4$&1.0\\
&$\beta_s g_v=-2\beta_Bg_v$&12.0\\
&$\lambda_sg_v=-2\sqrt{2}\lambda_Ig_v$&$19.2~\textrm{GeV}^{-1}$\\
&$g_\sigma$&$-3.65$\\
&$g$&12.00&\\
&$g_{VVP}$&$3g^2/(32\sqrt{2}\pi^2f_\pi)$\\
&$f_{\pi}$&0.132~$\textrm{GeV}$\\
\end{tabular}
\end{ruledtabular}
\end{table}

\subsection{Complex scaling method}\label{sec3}
At {present}, in order to obtain possible poles for these investigated systems, the complex scaling method (CSM) is applied. In the CSM, the relative distance $\bm{r}$ and the conjugate momentum $\bm{p}$ are replaced by
\begin{equation}
 \bm{r^{\prime}}\to\bm{r}e^{i\theta}, \quad \bm{p^{\prime}}\to\bm{p}e^{-i\theta}
\end{equation}
where the scaling angle $\theta$ is chosen to be { positive}. Applying such {a} replacement to the Schr\"odinger equation, we get the complex-scaled Schr\"odinger
equation for the coupled channels which {reads}
\begin{eqnarray}\label{S eq}
&&\left[\frac{1}{2\mu_{j}}\left(-\frac{d^{2}}{dr^{2}}+\frac{l_{j}(l_{j}+1)}{r^{2}}\right)e^{-2i\theta}+W_{j}\right]\psi_{j}^{\theta}(r) \nonumber \\ 
&& \qquad +\sum _{k}V_{jk}(re^{i\theta})\psi_{k}^{\theta}(r) =E\psi_{j}^{\theta}(r),
\end{eqnarray}
where $\mu_{j}$, $W_{j}$, and $\psi^{\theta}_j(r)$ are the reduced mass, corresponding threshold, and the orbital wave function, respectively.

It is worth {mentioning} that the properties of the solutions of the complex scaling Schr\"odinger
equation can be predicted by the so-called ABC theorem~\cite{Aguilar:1971ve,Balslev:1971vb}, which means
\begin{enumerate}
\item The wave functions for resonant states should be square-integrable on the complex plane, which is the same as {for a} bound state.
\item On the complex plane, the eigenvalues of the bound states and resonances are independent of the scaling angle $\theta$.
\item The continuum states change along the $2\theta$ line. 
\end{enumerate}
 \begin{figure}
      \centering
      \includegraphics[width=1\linewidth]{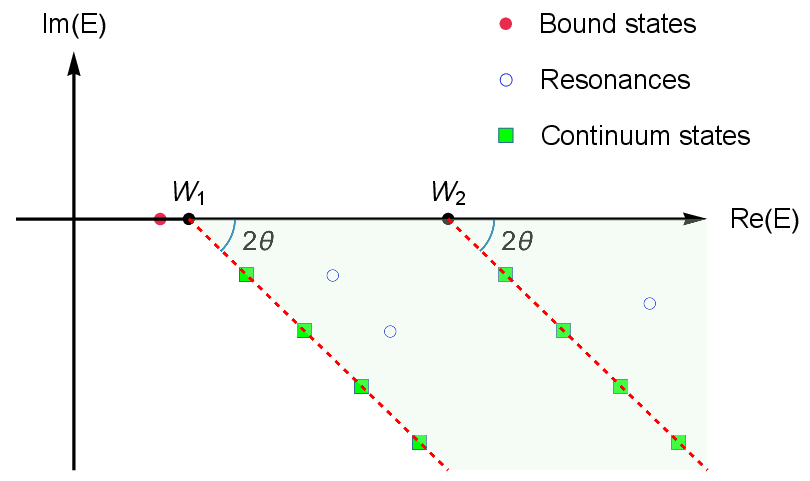}
      \caption{Schematic eigenvalue distributions of $H_\theta$ in the
coupled-channel two-body systems.}
      \label{csm}
  \end{figure}
 According to this theorem, {one pick out can the bound state or resonance from the continuum states as shown in Figure~\ref{csm}. This technique has been widely and successfully applied in quantum scattering theory and few-body systems. In our work, it provides a robust tool for identifying resonances in the coupled channel systems.}
Moreover, in this work, the orbital wave functions are expanded in terms of a set of Gaussian basis functions. With the obtained wave functions, the root-mean-square (RMS) radii $r_{RMS}$ and component proportions $P$ can be calculated by~\cite{T.yo,Lin:2023ihj}
\begin{subequations}
\begin{align}
	r_{RMS}^2&=\langle\psi^{\theta}|r^2|\psi^{\theta}\rangle=\sum_i\int r^2\psi^{\theta}_i(\bm{r})^2 d^3\bm{r},\\
	P&=\langle\psi^{\theta}_i|\psi^{\theta}_i\rangle=\int\psi^{\theta}_i(\bm{r})^2 d^3\bm{r}\label{eq:c-prod},
\end{align}
\end{subequations}
where the $\psi^{\theta}_i$ are normalized as
\begin{eqnarray}
	\sum_i\langle\psi^{\theta}_i\mid\psi^{\theta}_i\rangle=1.
\end{eqnarray}

{Interestingly, the root mean square (RMS) radius $r_{\rm RMS}$ for the resonances can be a complex number. In such cases, one can use the interpretation scheme proposed by T. Berggren, which generalizes the concept of expectation values from bound states to resonances~\cite{Berggren:1970wto}. According to this scheme, the real part of the complex $r_{\rm RMS}$ represents the usual physical expectation value, while the imaginary part indicates a measure of uncertainty in the observation. Numerical calculations of $r^2$ have supported this generalized interpretation~\cite{Gyarmati:1972yac, 1997matrix}. Besides, the $P$ involves the square of the complex-scaled wave function $\psi^{\theta}_i(\bm{r})$, rather than its modulus squared. Consequently, $P$ is not strictly positive definite. Depending on the nodal structure and the phase of the wave function, $P$ can take on positive or negative real values, or even have a complex part.}
{Moreover}, it is worth {mentioning} that the scaling angle $\theta$ should be larger than $1/2\text{Arg}(\Gamma/2E)$ to ensure the normalizability of {the} wave functions of the resonant states~\cite{Myo:2014ypa}.

\begin{figure*}
    \centering
    \includegraphics[scale=0.65]{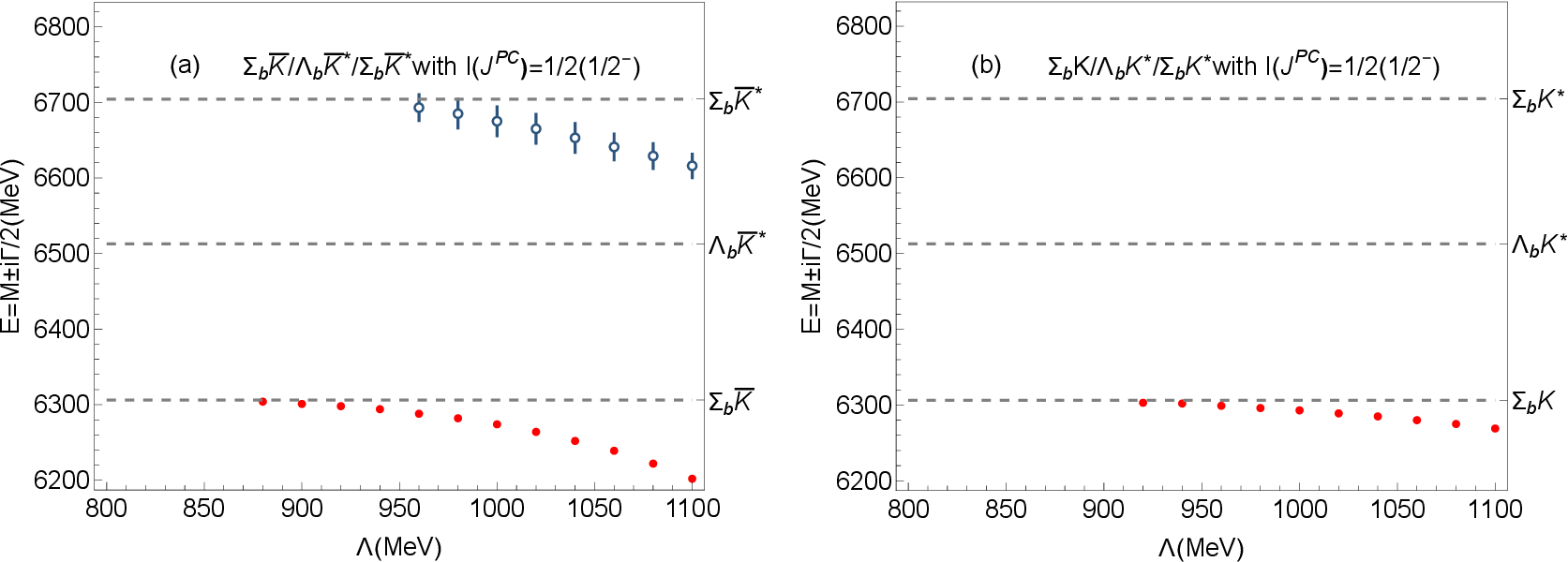}
    \caption{\label{massm}The $\Lambda$  dependence for the bottom-strange pentaquarks systems. The red solid dots stand for the bound states. The blue open circles with bars correspond to the resonances, with the lengths of bars being the total widths of the corresponding resonances.}
    \label{bar}
\end{figure*}

\section{Results and discussions}\label{results}

\begin{figure}
    \centering
    \includegraphics[width=1\linewidth]{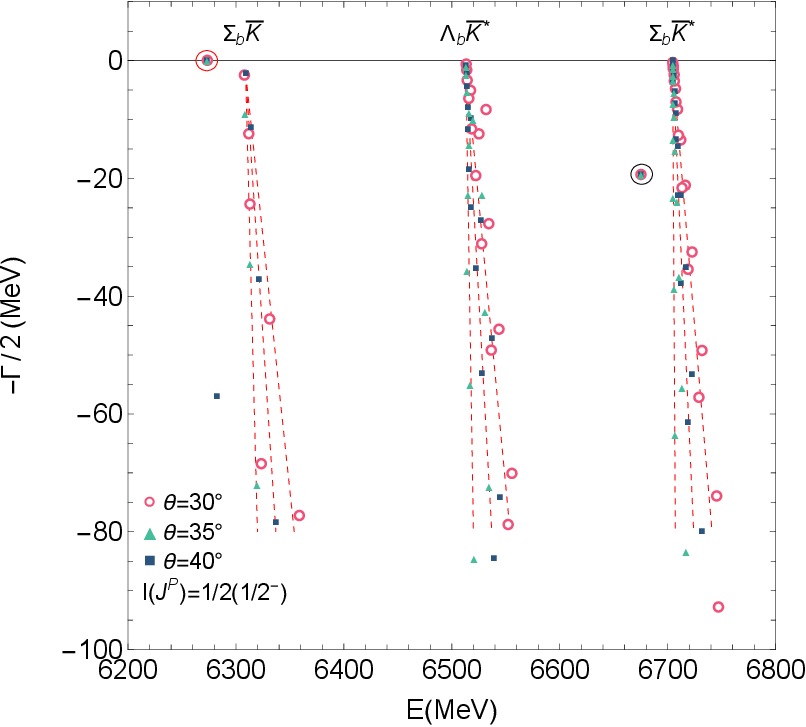}
   \caption{The complex energy eigenvalues of $I(J^P)=1/2(1/2^-)$ system by varying the angle $\theta$ from $30^\circ\sim40^\circ$.}
    \label{bc}
\end{figure}
\begin{figure}
    \centering
    \includegraphics[width=1\linewidth]{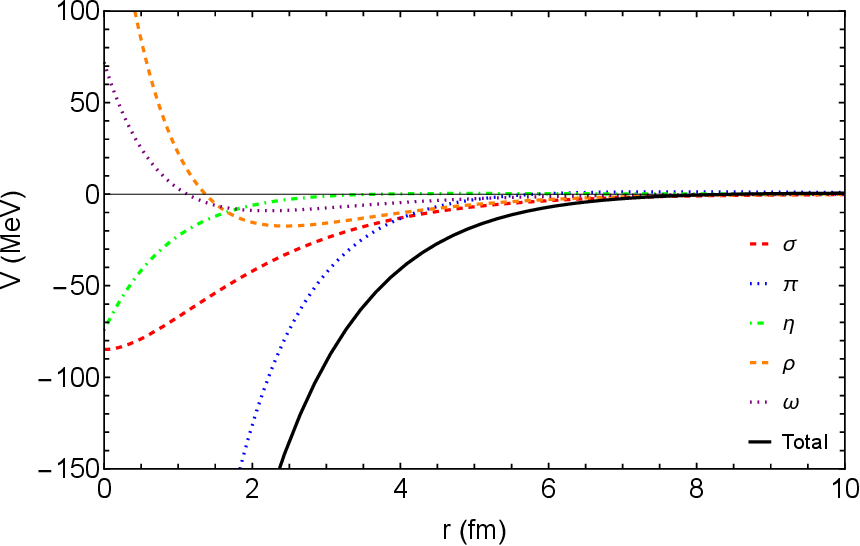}
    \caption{(Color online) The $r$ dependence of the deduced effective potentials with cutoff of 1000~$\rm{MeV}$ for the $S-$wave $\Sigma_b\bar{K}^*$ system with $I(J^P)=1/2(1/2^-)$.
}
    \label{fig_pt}
\end{figure}
Performing the above procedure, we can systematically investigate the bottom-strange molecular pentaquarks by solving {the} coupled channel Schr\"odinger equation. In this work, the only free parameter is the UV cutoff $\Lambda$ in Eq.~\eqref{FF},  which may vary for different coupled systems being investigated{,} and it {lies} within the range of $800\sim5000$ MeV~\cite{Liu:2009qhy,Chen:2019asm}. A reasonable cutoff for deuteron studies is estimated to be around 1000 $\rm MeV$, but it {can also} vary among different scenarios~\cite{Tornqvist:1993ng,Tornqvist:1993vu}. A loosely bound state at a cutoff around $1000$ $\rm MeV$ could be a promising molecular state candidate{;} thus, we summarize our predictions for bottom-strange pentaquark molecular state systems with cutoff $\Lambda$ in a range of $800\sim1100$ $\rm MeV$ in Table~\ref{sum}.

We firstly deal with {the} bottom baryon $Y_b$ and $\bar{K}^*$ meson systems to reveal possible bound and resonant states and give them reasonable interpretations. The same technique can be applied in the analysis of {the} bottom baryon $Y_b$ and $K$ meson systems. Our estimations for these investigated systems depending on the cutoff value $\Lambda$ are plotted in Figure~\ref{massm} and listed in Table~\ref{zb} and~\ref{re}. In the present work, both $S$-$D$ wave mixing effects and coupled channel effects are taken into account. According to the isospin, spin, and parity, the bottom baryon and anti-strange meson systems can be classified as $1/2(1/2^-)\Sigma_{b}\bar{K}/\Lambda_{b}\bar{K}^*/\Sigma_b\bar{K}^*$, $3/2(1/2^-)\Sigma_b\bar{K}/\Sigma_b\bar{K}^*$, $1/2(3/2^-)\Lambda_{b}\bar{K}^*/\Sigma_b\bar{K}^*$, and $3/2(3/2^-)\Sigma_b\bar{K}^*$ channels, respectively. The corresponding classification also exists for the bottom baryon and strange meson systems.

\begin{table*}[!t]
	\renewcommand\arraystretch{1.4}
	\caption{\label{zb} The numerical results for the obtained bound states.}
	\begin{ruledtabular}
		\begin{tabular}{ccccccccccc}
        		&&$\Lambda(\rm{MeV})$
			&$r_{RMS}(\rm{fm})$
			&$E(\rm{MeV})$&$(\Sigma_b\bar{K}(^{2}S_{1/2})$&$\Sigma_b\bar{K}^*(^{2}S_{1/2})$&$\Sigma_b\bar{K}^{*}(^{4}D_{3/2}))$
			\\\hline
   \multirow{3}{*}{$I(J^P)=3/2(1/2^-)$}
			&&3600&3.77&6304.84& (99.08&0.47&0.45)\\
  & &3850&1.44&6292.68&(95.75&2.28&1.97)\\
   &&4100&1.03&6278.37& (92.78&3.97&3.25)\\
   \hline
			&&$\Lambda(\rm{MeV})$
			&$r_{RMS}(\rm{fm})$
			&$E(\rm{MeV})$&($\Lambda_b\bar{K}^{*}(^{4}S_{3/2})$&$\Lambda_b\bar{K}^{*}(^{2}D_{3/2})$&$\Lambda_b\bar{K}^{*}(^{4}D_{3/2})$&$\Sigma_b\bar{K}^{*}(^{4}S_{3/2})$&$\Sigma_b\bar{K}^{*}(^{2}D_{3/2})$&$\Sigma_b\bar{K}^{*}(^{4}D_{3/2})$)
			\\\hline
\multirow{3}{*}{{$I(J^P)=1/2(3/2^-)$}} 
  &&{1460}&{1.06}&{6511}& 
  ({12.19}&{0.52}&{2.98}&{75.64}&{1.15}&{7.52})\\
  &&{1480}&{0.59}&{6493}&
  ({4.23}&{0.53}&{3.00}&{82.90}&{1.22}&{8.12})\\
  &&{1500}&{0.55}&{6474}&
  ({3.14}&{0.50}&{2.88}&{83.93}&{1.23}&{8.32})\\
   \hline
   &&$\Lambda(\rm{MeV})$
			&$r_{RMS}(\rm{fm})$
			&$E(\rm{MeV})$&($\Sigma_b\bar{K}^{*}(^{4}S_{3/2})$&$\Sigma_b\bar{K}^{*}(^{2}D_{3/2})$&$\Sigma_b\bar{K}^{*}(^{4}D_{3/2})$)
			\\\hline
   \multirow{3}{*}{$I(J^P)=3/2(3/2^-)$}
      &&1300&3.90&6704&(99.26&0.16&0.58)\\
   &&1400&2.36&6721&(98.67&0.29&1.04)\\
   &&1500&1.61&6697&(98.17&0.40&1.43)\\\hline
   &&$\Lambda(\rm{MeV})$
			&$r_{RMS}(\rm{fm})$
			&$E(\rm{MeV})$&$(\Sigma_bK(^{2}S_{1/2})$&$\Sigma_bK^*(^{2}S_{1/2})$&$\Sigma_bK^{*}(^{4}D_{1/2}))$
			\\\hline
   \multirow{3}{*}{$I(J^P)=3/2(1/2^-)$}
			&&1260&1.51&6298.01& (76.17&23.72&0.10)\\
  & &1270&0.90&6284.01&(65.82&34.04&0.15)\\
   &&1280&0.67&6265.52& (58.84&41.00&0.16)\\
   \hline
   	&&$\Lambda(\rm{MeV})$
			&$r_{RMS}(\rm{fm})$
			&$E(\rm{MeV})$&($\Lambda_bK^{*}(^{4}S_{3/2})$&$\Lambda_bK^{*}(^{2}D_{3/2})$&$\Lambda_bK^{*}(^{4}D_{3/2})$&$\Sigma_bK^{*}(^{4}S_{3/2})$&$\Sigma_bK^{*}(^{2}D_{3/2})$&$\Sigma_bK^{*}(^{4}D_{3/2})$)
			\\\hline
\multirow{3}{*}{{$I(J^P)=1/2(3/2^-)$}} 
  &&{1320}&{2.02}&{6510}& 
  ({70.16}&{0.09}&{0.37}&{27.76}&{0.32}&{1.30})\\
  &&{1340}&{0.96}&{6502}&
  ({52.39}&{0.11}&{0.54}&{44.70}&{0.43}&{1.83})\\
  &&{1360}&{0.70}&{6490}&
  ({43.51}&{0.12}&{0.58}&{53.39}&{0.45}&{1.95})

		\end{tabular}
	\end{ruledtabular}
\end{table*}

\begin{table*}[!t]
	\renewcommand\arraystretch{1.4}
	\caption{\label{re} The numerical results for the bottom baryon and anti-meson interactions with $I(J^P)=1/2(1/2^-)$ .}
	\begin{ruledtabular}
		\begin{tabular}{ccccccccccc}
        		&&$\Lambda(\rm{MeV})$
			&$r_{RMS}(\rm{fm})$
			&$E(\rm{MeV})$&($(\Sigma_b\bar{K}^*(^{2}S_{1/2})$&$\Sigma_b\bar{K}^{*}(^{4}D_{1/2})$)
			\\\hline
   \multirow{3}{*}{$I(J^P)=1/2(1/2^-)$}
			&&840&4.37&6703.34& (99.65&0.35)\\
  & &970&1.16&6687.43&(99.15&0.85)\\
   &&1100&0.78&6651.48& (98.83&1.17)\\
   \hline
   &&$\Lambda(\rm{MeV})$
			&$r_{RMS}(\rm{fm})$
			&$E(\rm{MeV})$&($(\Sigma_b\bar K(^{2}S_{1/2})$&$\Sigma_b\bar K^*(^{2}S_{1/2})$&$\Sigma_b\bar K^{*}(^{4}D_{1/2})$)
			\\\hline
   \multirow{3}{*}{$I(J^P)=1/2(1/2^-)$}
			&&900&{$1.89-0.20i$}&{$6699.72-0.96i$}& ({$-0.15-0.13i$}&{$99.39 +0.13 i$}&{$0.76 -0.00 i$})\\
  & &1000&{$1.12 +0.00i$}&{$6680.00-0.07i$}&({$0.02 +0.04 i$}&{$99.08 -0.05 i$}&{$0.90 +0.01i$})\\
   &&1100&{$0.82-0.03i$}&{$6643.78-12.99i$}& ({$-1.13+0.29i$}&{$100.64-0.62i$}&{$0.49 +0.33i$})\\
   \hline
			&&$\Lambda(\rm{MeV})$
			&$r_{RMS}(\rm{fm})$
			&$E(\rm{MeV})$&(($\Lambda_b\bar{K}^{*}(^{2}S_{1/2})$&$\Lambda_b\bar{K}^{*}(^{4}D_{1/2})$&$\Sigma_b\bar{K}^{*}(^{2}S_{1/2})$&$\Sigma_b\bar{K}^{*}(^{4}D_{1/2})$)
			\\\hline
   \multirow{3}{*}{$I(J^P)=1/2(1/2^-)$}
						&&920&{$2.46 +5.02 i$}&{$6697.48-33.62i$}& ({$-16.57+2.89 i$}&{$0.28 +1.26 i$}&{$116.26 -4.14 i$}&{$0.03-0.01i$})\\
  & &1010&{$1.01 -0.22i$}&{$6651.10-42.66i$}&({$-31.38+10.64i$}&{$0.95+1.67i$}&{$130.03-13.15i$}&{$0.40 +0.84i$})\\
   &&1100&{$0.88 -0.27i$}&{$6605.03-31.83i$}& ({$-13.67+28.50i$}&{$1.46 +0.43i$}&{$110.82 -29.83 i$}&{$1.39+0.90i$})\\
   \hline
   &&$\Lambda(\rm{MeV})$
			&$r_{RMS}(\rm{fm})$
			&$E(\rm{MeV})$&($\Sigma_b\bar K(^{2}S_{1/2})$&$\Lambda_b\bar{K}^{*}(^{2}S_{1/2})$&$\Lambda_b\bar{K}^{*}(^{4}D_{1/2})$&$\Sigma_b\bar{K}^{*}(^{2}S_{1/2})$&$\Sigma_b\bar{K}^{*}(^{4}D_{1/2})$)
			\\\hline
   \multirow{3}{*}{$I(J^P)=1/2(1/2^-)$}
						&&960&{$1.22-0.55 i$}&{$6692.84-17.29i$}& ({$1.24 +0.25 i$}&{$-2.99-4.37 i$}&{$-1.15+0.68i$}&{$102.41 +3.33 i$}&{$0.49 +0.11i$})\\
  & &1030&{$0.98 -0.09i$}&{$6659.05-18.99i$}&({$1.36 +0.35 i$}&{$-5.99+0.17i$}&{$-0.46+1.34i$}&{$104.28 -2.40 i$}&{$0.81 +0.54i$})\\
   &&1100&{$0.84 -0.10i$}&{$6615.80-15.50i$}& ({$0.89 +0.51 i$}&{$-4.14+7.39i$}&{$0.40 +1.11i$}&{$101.57-9.75i$}&{$1.28 +0.74i$})\\
		\end{tabular}
	\end{ruledtabular}
\end{table*}

\subsection{Bottom baryon and anti-strange meson systems}
\subsubsection{{$I(J^{P})=1/2(1/2^-)$ with $\Sigma_{b}\bar{K}/\Lambda_{b}\bar{K}^*/\Sigma_b\bar{K}^*$} }

In this subsection, we firstly discuss the coupled $I(J^{PC})=1/2(1/2^{-})$ $\Sigma_{b}\bar{K}/\Lambda_{b}\bar{K}^*/\Sigma_b\bar{K}^*$ systems. As shown in Fig.~\ref{massm}(a), one could find that a bound state and a resonance emerge within the range $\Lambda = 880\sim1100$ $\rm{MeV}$. When the cutoff $\Lambda$ is set to 880 $\rm{MeV}$, the bound state {is located} below the $\Sigma_b\bar{K}$ threshold with {a} binding energy {of about} $4$ $\rm{MeV}$, the $r_{RMS}$ is 3 $\rm{fm}$, and {it is} dominated by the $\Sigma_b\bar{K}(^2S_{1/2})$ channel. Sliding the cutoff to $1080$ $\rm{MeV}$, the mass {becomes} around 6222 $\rm{MeV}$ and the $r_{RMS}$ {is} 0.8 $\rm{fm}$, which is consistent with the sizes of exotic hadronic molecular {states}. Thus, this bound state is a good candidate {for} the particle $\Xi_b(6227)$. These results {favor} the conclusion in Refs.~\cite{Huang:2018bed,Zhu:2020lza,Mutuk:2024elj}. 

Meanwhile, { at a cutoff of $\Lambda = 960~\rm{MeV}$, we obtain a quasi-bound state with complex energy $E = 6693 - 17i~\rm{MeV}$ and root-mean-square radius $r_{\rm RMS} = 1.22-0.55 i~\rm{fm}$, which is predominantly composed of the $\Sigma_b\bar{K}^*(^2S_{1/2})$ channel. The binding of this state mainly arises from the strong interaction in the (closed) $\Sigma_b\bar{K}^*$ channel, while the nonzero width is induced via its coupling to lower open channels such as $\Lambda_b \bar{K}^*$. This structure is consistent with the nature of a Feshbach-type resonance and can be regarded as a good hadronic molecular candidate.}

{Besides, We illustrate the complex energy eigenvalues of the $I(J^P)=1/2(1/2^-)$ system while varying the angle $\theta$ from $30^\circ \sim 40^\circ$ in Figure~\ref{bc}.
Specifically, the complex energy plane reveals the nature of poles: bound states lie on the real axis below threshold, while resonance appears as pole in the complex plane. The behavior of continuum states under varying scaling angle $\theta$ illustrates the separation between physical and unphysical states. 
Additionally, the clustering of poles near certain thresholds, such as $\Sigma_b \bar K$, $\Lambda_b \bar K^*$, and $\Sigma_b \bar K^*$ highlights the role of channel coupling and threshold effects. These patterns help identify the dominant configurations and offer insight into the underlying interaction dynamics.}

{Furthermore, our results highlight that the pion-exchange potential plays a crucial role in forming this resonance, whereas the contribution from the $\eta$-exchange interaction is negligible. As shown in Fig.~\ref{fig_pt}, the $\eta$ exchange provides a much weaker attractive potential compared to the $\pi$ exchange. The dominant role of the pion exchange arises from its lighter mass and longer range, which enhance the attractive interaction between hadrons. In contrast, the $\eta$ meson is heavier, and its contribution is significantly suppressed. This physical picture is now clearly stated in the revised text. Moreover, according to our calculations, the resonance pole still appears even when the $\eta$-exchange potential is omitted, provided that the same cutoff is used.}

\subsubsection{{$I(J^{P})=3/2(1/2^-)$ with $\Sigma_{b}\bar{K}/\Sigma_b\bar{K}^*$ }}
{For the $I(J^{P})=3/2(1/2^-)\Sigma_{b}\bar{K}/\Sigma_b\bar{K}^*$ state, when the cutoff $\Lambda$ lies in the range {from} $3360$ {to} $4100$ $\rm{MeV}$, a loosely bound state mainly composed of the $\Sigma_b\bar{K}(^2S_{1/2})$ component (about 93\%$\sim$99\%) is found, with its mass varying from 6304.8 to 6278~$\rm{MeV}$ and the root-mean-square radius $r_{\rm RMS}$ ranging from 4 to 1~$\rm{fm}$.
 However, this {cutoff range} is quite different from the empirical value for the deuteron, and {thus} no molecular state is favored in the $3/2(1/2^-)$ $\Sigma_b\bar{K}/\Sigma_b\bar{K}^*$ system.}

 \subsubsection{{{$I(J^{P})=1/2(3/2^-)$ with $\Lambda_{b}\bar{K}^*/\Sigma_b\bar{K}^*$ }}}
{
Besides, we investigate the $I(J^{P})=1/2(3/2^-)$ $\Lambda_{b}\bar{K}^{*}/\Sigma_{b}\bar{K}^{*}$ channel for the $Y_b$ and $\bar{K}^{(*)}$ system, and the corresponding results are listed in Table~\ref{zb}. According to our estimations, at a cutoff of 1460~MeV, a bound state appears below the $\Lambda_b\bar{K}^*$ threshold. This state is mainly composed of the $\Sigma_b\bar{K}^*(^4S_{3/2})$ (75.6\%), $\Sigma_b\bar{K}^*(^4D_{3/2})$ (7.5\%), and $\Lambda_b\bar{K}^*(^4S_{3/2})$ (12\%) components, and shows a strong dependence on the cutoff value. However, since this cutoff is inconsistent with the empirical value used for describing deuteron systems, and the corresponding $r_{\rm RMS}$ does not fall within the typical range for a molecular state, we conclude that the $I(J^{P})=1/2(3/2^-)$ $\Lambda_{b}\bar{K}^{*}/\Sigma_{b}\bar{K}^{*}$ system is unlikely to form a hadronic molecular state.
}

 \subsubsection{{$I(J^{P})=3/2(3/2^-)$ with $\Sigma_b\bar{K}^*$ }}
In the $I(J^{P})=3/2(3/2^-)$ $\Sigma_b\bar{K}^*$ system, a weakly bound state with {an} energy of $6703$ $\rm{MeV}$ appears at {a} cutoff of $1300$ $\rm{MeV}$ and is dominated by {the $S$-wave channel about 99\%}. If only the {one-pion-exchange} potential is considered, a bound state is {obtained} with {a} cutoff of $2400$ $\rm{MeV}$, which means that the potentials of $\rho$ and $\omega$ exchanges are helpful to form the bound state.

\subsection{Bottom baryon and strange meson systems}
For $Y_bK^{(*)}$ systems, the effective potentials from the $\omega$ and $\pi$ exchanges are {in complete contrast} with {those of the} $Y_b\bar{K}^{(*)}$ systems. Unlike {the} bottom baryon and anti-strange meson systems, {for the} bottom baryon and strange meson systems, one can only obtain bound state solutions. The corresponding numerical results are collected in Table~\ref{zb} and Figure~\ref{massm}(b).

 \subsubsection{{$I(J^{P})=1/2(1/2^-)$ with $\Sigma_b K / \Lambda_b K^* / \Sigma_b K^*$ }}
 {
    When the cutoff parameter varies within $920 \sim 1100~\rm{MeV}$, the mass decreases from 6304 to 6269~$\rm{MeV}$, while the root-mean-square radius $r_{RMS}$ shrinks from 3 to 1~$\rm{fm}$, consistent with expectations for a molecular candidate. Considering only the single $\Sigma_b K$ channel, a bound state solution also emerges at a larger cutoff of $\Lambda = 1900~\rm{MeV}$, further indicating that coupled-channel effects play a significant role in forming the molecule.}

 \subsubsection{{$I(J^{P})=3/2(1/2^-)$ with $\Sigma_b K / \Sigma_b K^*$}}
 For the $I(J^{P}) = 3/2(1/2^-)$ $\Sigma_b K / \Sigma_b K^*$ system, a bound state solution is found {when the cutoff lies in the $1260~\rm MeV$, which is mainly composed of $S-$ wave $\Sigma_b K$ and $\Sigma_b K^*$ about $76\%$ and 24\%, respectively. When we vary the cutoff to $1270~\rm MeV$, the predicted mass varies from 6298 to 6284~$\rm{MeV}$, with the corresponding $r_{RMS}$ decreasing from $1.5$ to $0.9~\rm{fm}$. This sensitivity to the cutoff suggests the possibility of a molecular state.}
 \subsubsection{{$I(J^P) = 1/2(3/2^-)$ with $\Lambda_b K^* / \Sigma_b K^*$ }}
{As listed in Table~\ref{zb}, for the $\Lambda_b K^* / \Sigma_b K^*$ system with $I(J^P) = 1/2(3/2^-)$, a bound state below the $\Lambda_b K^*$ threshold appears at a cutoff of 1260~$\rm{MeV}$. However, if only the $\Lambda_b K^*$ channel is considered, no bound state pole is found for cutoff values ranging from 800 to 5000~$\rm MeV$. In contrast, a loosely bound state dominated by the $\Sigma_b K^*(^4S_{3/2})$ component (about 98\%) appears when considering only the $\Sigma_b K^*$ channel, with a cutoff of 960~$\rm MeV$. This state has a mass of 6704~$\rm MeV$ and a root-mean-square radius $r_{\rm RMS} = 4~\rm fm$.
}
 \subsubsection{{$I(J^P) = 3/2(3/2^-)$ with $\Sigma_b K^*$ }}
Finally, for the $I(J^P) = 3/2(3/2^-)$ $\Sigma_b K^*$ system, no bound state solution is found within the cutoff range $\Lambda = 800 \sim 5000~\rm{MeV}$.

   \begin{table*}[!htbp]
 	\renewcommand\arraystretch{1.4}
 	\caption{\label{sum} The summary of our predictions for bottom-strange  pentaquark molecular state  systems with cutoff $\Lambda$ in a range of $800\sim1100$ $\rm MeV$. Here, the , $"\checkmark"("\times")$ represents that the corresponding state may (may not) form a molecular state.}
 	\begin{ruledtabular}
 		\begin{tabular}{cccccccc}
 			&$I(J^{PC})$
 			&Mass$(\rm MeV)$&Width$(\rm MeV)$&$r_{RMS}(\rm fm)$&Status& Selected decay mode
 			\\\hline
 			&\multirow{2}{*}{$\frac{1}{2}(\frac{1}{2}^{-})$$\Sigma_b\bar{K}/\Lambda_b\bar{K}^*/\Sigma_b\bar{K}^*$}&$6301\sim6222$&$-$&$2.49\sim0.73$&$\checkmark$&$\Lambda_b\bar{K}/\Xi_{b}^{(\prime)}\pi$\\
&&$6693\sim6516$&$34.58\sim31.00$&{$1.22-0.55i\sim0.84- 0.10i$}&$\checkmark$&$\Lambda_b\bar{K}^{(*)}/\Sigma_b\bar{K}/\Lambda\bar{B}^{(*)}/\Sigma\bar{B}/\Xi_{b}^{(\prime)}\pi/\Xi_b^{(\prime)}\eta/\Xi_{b}\rho/\Xi_b\omega$\\\hline
&\multirow{1}{*}{$\frac{3}{2}(\frac{1}{2}^{-})$$\Sigma_b\bar{K}/\Sigma_b\bar{K}^*$}&$-$&$-$&$-$&$\times$&$-$\\\hline

&$\frac{1}{2}(\frac{3}{2}^{-})$$\Lambda_b\bar{K}^*/\Sigma_b\bar{K}^*$&$-$&$-$&$-$&$\times$&$-$\\\hline
&$\frac{3}{2}(\frac{3}{2}^-)\Sigma_b\bar{K}^*$&$6703\sim6702$&$-$&$3.37\sim2.62$&$\checkmark$&$\Lambda_b\bar{K}^{*}/\Sigma_b^*\bar{K}/\Lambda\bar{B}^{*}/\Sigma\bar{B}^*/\Xi_{b}^{*}\pi/\Xi_b^{*}\eta/\Xi_{b}\rho/\Xi_b\omega$\\
\hline
&$\frac{1}{2}(\frac{1}{2}^{-})$$\Sigma_bK/\Lambda_bK^*/\Sigma_bK^*$&$6303\sim6269$&$-$&$3.13\sim1.08$&$\checkmark$&$N\bar{B}_s/\Lambda_bK$\\\hline
&\multirow{1}{*}{$\frac{3}{2}(\frac{1}{2}^{-})$$\Sigma_bK/\Sigma_bK^*$}&$-$&$-$&$-$&$\times$&$-$\\\hline
&$\frac{1}{2}(\frac{3}{2}^{-})$$\Lambda_bK^*/\Sigma_bK^*$&$-$&$-$&$-$&$\times$&$-$\\\hline
&$\frac{3}{2}(\frac{3}{2}^{-})$$\Sigma_bK^*$&$6704\sim6692$&$-$&$4.13\sim1.34$&$\checkmark$&$N\bar{B}_s^{*}/\Lambda_b^*K/\Sigma_b^*K$\\
 		\end{tabular}
 	\end{ruledtabular}
 \end{table*}


\subsection{Further discussions}

For $Y_b\bar{K}^{(*)}$ systems, we can obtain bound states and resonances, but only bound states are revealed {for} $Y_bK^{(*)}$ systems. The reason {is that the flavor factors in the potentials for these systems differ significantly}, which {determine} the relative sign and strength and are crucial for the formation of molecular states.

In our study, we aim to explore the crucial role of coupled-channel effects in the production of resonant states. As shown in Table~\ref{re}, when considering only the single-channel $\Sigma_b\bar{K}^{*}$ $S$-$D$ wave mixing, we obtain a bound state at a cutoff of 840 $\rm{MeV}$. However, when including the coupled-channel effect for the $I(J^{P})=1/2(1/2^-)$ $\Sigma_b\bar{K}/\Sigma_b\bar{K}^{*}$ system, we find a relatively narrow resonant state at a cutoff of 900 $\rm{MeV}$, which is dominated by the $I(J^{P})=1/2(1/2^-)$ $\Sigma_b\bar{K}^{*}(^{2}S_{1/2})$ component. Additionally, for the coupled-channel system $I(J^{P})=1/2(1/2^-)$ $\Lambda_b\bar{K}^{*}/\Sigma_b\bar{K}^{*}$, we obtain a relatively broad resonance at a cutoff of 920 $\rm{MeV}$. Furthermore, when considering the fully coupled $I(J^{P})=1/2(1/2^-)$ $\Sigma_b\bar{K}/\Lambda_b\bar{K}^{*}/\Sigma_b\bar{K}^{*}$ system, a resonance emerges near the $\Sigma_b\bar{K}^{*}$ threshold, with a complex energy 
$E {-} i\Gamma/2 = 6692.84 {-} 17.29i$ at a cutoff of 960 $\rm{MeV}$. On the other hand, when we discard the $D$-wave contributions, we still obtain a resonant state at a cutoff of approximately 1000 $\rm{MeV}$ for the $S$-wave coupled systems $\Sigma_b\bar{K}/\Sigma_b\bar{K}^{*}$, $\Lambda_b\bar{K}{/}\Sigma_b\bar{K}^{*}$, and $\Sigma_b\bar{K}/\Lambda_b\bar{K}^{*}/\Sigma_b\bar{K}^{*}$ with $I(J^P)=1/2(1/2^-)$, respectively. In conclusion, our calculations demonstrate that the coupled-channel effect plays a significant role in the formation of resonant states, with the main contributions coming from the $S$-wave.

According to the masses and quantum numbers, we present some possible decay channels for these predicted states in Table~\ref{sum}. For instance, the $I(J^{PC})=1/2(1/2^{-})$ $\Sigma_{b}\bar{K}/\Lambda_{b}\bar{K}^*/\Sigma_b\bar{K}^*$ bound state can be found in the $\Lambda_b\bar{K}$ and $\Xi_{b}^{(\prime)}\pi$ channels. In the literature~\cite{Chen:2018orb,He:2021xrh,Wang:2018fjm,Cui:2019dzj}, both $J^P=\frac{1}{2}^-$ molecular and $J^P=\frac{3}{2}^-/\frac{5}{2}^-$ conventional interpretations exist for the particle $\Xi_b(6227)$, and the spin is certainly crucial for distinguishing these two explanations. Another way to solve this puzzle is to hunt for the flavor exotic state with mass $6303\sim6269~\rm{MeV}$ in the $I(J^{PC})=1/2(1/2^{-})$ $\Sigma_{b}K/\Lambda_{b}K^*/\Sigma_bK^*$ system, which is the mirror state of $\Xi_b(6227)$ in the molecular picture but does not appear in the three-quark picture. We {strongly} hope that future experiments can verify our proposals.

\section{Summary}\label{sec4}

In this work, we systematically investigate the coupled $Y_b\bar{K}^{(*)}(Y_bK^{(*)})$ system to search for possible bound states and resonances by adopting {the} one-boson-exchange model within {the} complex scaling method. For the coupled $I(J^P)=1/2(1/2^-)$ $\Sigma_b\bar{K}/\Lambda_b\bar{K}^*/\Sigma_b\bar{K}^*$ systems, according to our estimations, a bound state solution is obtained, which may correspond to the observed particle $\Xi_b(6227)$. Meanwhile, we find a $I(J^{PC})=1/2(1/2^-)$ resonance near the $\Sigma_b\bar{K}^*$ threshold and a bound state in the $I(J^P)=1/2(3/2^-)$ $\Sigma_b\bar{K}^*$ system.

Then, when we extend our study to the $Y_{b}K^{(*)}$ systems, two loosely bound states are obtained. It is worth pointing out that the predicted bound state with mass $6303\sim6269~\rm{MeV}$ in the $I(J^P)=1/2(1/2^-)$ $\Sigma_bK/\Lambda_bK^*/\Sigma_bK^*$ system is flavor exotic and does not appear in the spectroscopy of conventional baryons, which provides a practical way to resolve the puzzle of the particle $\Xi_b(6227)$. We hope our predictions can offer valuable information {to future experimental} observations.

\subsection*{ACKNOWLEDGMENTS}

We would like to thank Rui Chen, Xian-Hui Zhong,  Li-Cheng Sheng, and Jin-Yu Huo for useful discussions. The work of X.-N. X. and Q. F. Song is supported by the National Natural Science
Foundation of China under Grants No. 12275364.  Q.-F. Lü is supported by the Natural Science Foundation of Hunan Province under Grant No. 2023JJ40421, the Scientific Research Foundation of Hunan Provincial Education Department under Grant No. 24B0063, and the Youth Talent Support Program of Hunan Normal University under Grant No. 2024QNTJ14.

\end{document}